\documentclass[9pt,twoside]{revtex4-1}

\usepackage{hyperref}
\usepackage{amsmath,amssymb,graphicx,epstopdf}

\begin{document}

\title{Continuous generation of Rubidium vapor in hollow-core photonic band-gap fibers}

\author{Prathamesh S. Donvalkar}
\email{psd52@cornell.edu}
\author{Sven Ramelow}
\author{St{\'e}phane Clemmen}
\author{Alexander L. Gaeta}
\affiliation{School of Applied and Engineering Physics, Cornell University, Ithaca, NY 14853}

\begin{abstract}
We demonstrate high optical depths (50${\pm}$5), lasting for hours in Rubidium-filled hollow-core photonic band-gap fibers, which represents a 1000X improvement over operation times previously reported. We investigate the vapor generation mechanism using both a continuous-wave and a pulsed light source and find that the mechanism for generating the Rubidium atoms is primarily due to thermal vaporization. Continuous generation of large vapor densities should enable measurements at the single-photon level by averaging over longer time scales.
\end{abstract}

\maketitle

Nonlinear interactions between photons is a longstanding goal in quantum optics. Large photon-photon interaction strengths have been realized by means of creating a Rydberg blockade in cold atomic ensembles \cite{Payeronel} or by using cavity quantum electrodynamics \cite{Chen,Tiecke,Ritter,Rauschenbeutel}. Such ultralow power light matter interactions can be applied to quantum information processing \cite{Sven}, quantum non-demolition measurements by the process of cross-phase modulation \cite{Imamoglu}, and noiseless frequency conversion \cite{Kumar}. A less complex approach for strong photon-photon interactions is to use systems that exhibit large Kerr nonlinearities. The requirements for achieving such nonlinearities in atomic/molecular vapors are a high optical depth (OD) and  small cross sectional area to achieve high intensities for low power levels \cite{Bhagwat}. Alkali metal vapors confined to hollow-core photonic band-gap fibers (PBGF) have  extremely large optical nonlinearities \cite{Saikat,Pablo,Viv1,Sprague,Epple}, and large cross phase shifts \cite{Vivek_natphoton, Luiten}, all optical amplitude modulation \cite{Vivek_prl}, frequency translation \cite{Prathamesh} and  quantum memory \cite{Walmsley} have been demonstrated using such systems. More recently, high OD's have been observed by loading cold atoms in a PBGF from a MOT \cite{Blatt} and the diffusive loading of a kagom\'e style PBGF with mercury vapor, capable of achieving high optical nonlinearities in the UV regime \cite{russel_hg}. A drawback for the warm Rb-filled PBGF platform has been the short time scales of operation, ranging from about a few seconds to a minute. Furthermore, the system requires an extended duration ($> 3$ hrs) to re-establish high OD operation \cite{Bhagwat,Sprague,slepkov_pra}. Theories have been proposed to understand the dynamics of Rb atoms inside the fiber core and their interaction with the inner walls of the core \cite{Bhagwat}. A detailed analysis suggests that Rb nanoclusters form inside the core of the fibers, and their subsequent vaporization using intense beams is responsible for the high OD. However, the Rb vapor obtained by thermal vapor generation from the nano-clusters is quickly lost to the fiber walls. Thus, high OD's that are a key to achieving high third order nonlinearities are available only for a few seconds which largely limits the operation time of the system.

\begin{figure}[ht]
\centerline{\includegraphics[width=0.65\columnwidth]{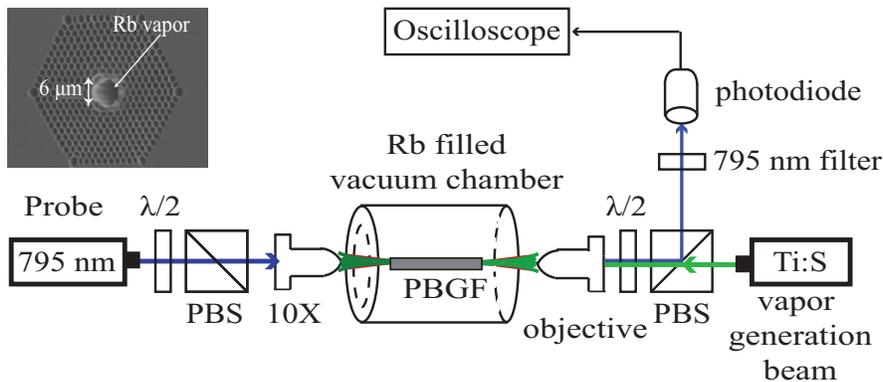}}
\caption{A weak probe beam at 795 nm and a vapor generation beam are sent counterpropagating into the Rb-filled photonic band-gap fiber (PBGF) using 10X objectives. The probe is scanned across the $D_{1}$ line of Rb. After propagating through the fiber, the probe beam is cleared out using spectral filters from any stray light that originates from the back reflected component of the vapor generation beam. The absorption of the probe beam is measured with a photodiode.}
\end{figure}

In this work, we report a 1000X improvement in interaction times in Rb-PBGF over  those previously achieved in such systems for OD's ${\textgreater}$ 50. Such long operating times were achieved using vapor generation beams with substantially higher average powers and wavelengths close to the band edge of the PBGF. The generated OD's are commensurate with those required for all previous demonstrations of optical modulation using the Rb-PBGF system \cite{Pablo, Viv1, Vivek_natphoton, Vivek_prl, Prathamesh, Kasturi}. This capability greatly facilitates performing experiments related to weak nonlinearity-based quantum computing \cite{Munro}. Varying the power and wavelength of the vapor generation beam offers further insight into the mechanism leading to sustained OD's. We also investigate vapor generation using mode-locked pulses and compare it with our results using a continuous-wave (CW) beam, and we conclude that it is the average and not the peak power which facilitates the generation of high OD's. 

\begin{figure}[ht]
\centerline{\includegraphics[width=0.5\columnwidth]{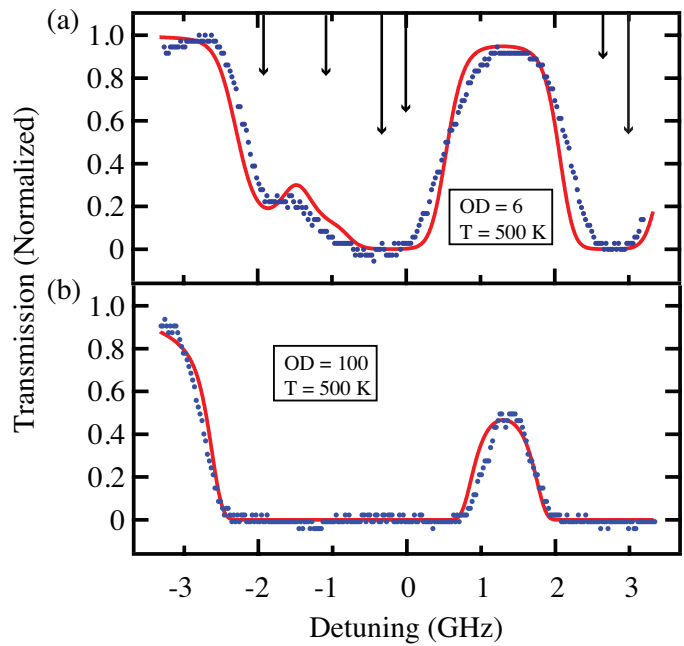}}
\caption{The absorption of the probe beam is plotted as a function of detuning, for vapor generation beam at (a) 17.3 mW, (b) 48 mW. The frequency scale is zeroed at the F = 3 $\rightarrow$ F' = 3 transition on the $^{85}$Rb $D_{1}$ line. Each of the above absorption curves is fit to a Voigt function taking into account the Doppler and transit time broadened hyperfine transition on the $^{85}$Rb and $^{87}$Rb $D_{1}$ line. From the fit, the temperature of the Rb vapor in the core can be estimated to be $T$ $\sim$ 500 K.}
\end{figure}

The Rb chamber design containing the PBGF is the same as that used previously                 \cite{Vivek_natphoton,Vivek_prl,Prathamesh, Kasturi}. A standard cleaning procedure for vacuum components was followed, after which we pre-baked the chamber for over a month to achieve ultrahigh vacuum ($< 10^{-9}$ Torr) before the Rb vapor was released. The Rb density slowly builds up in the core of the PBGF over a period of two weeks by diffusion of warm Rb vapor from the chamber.  A 9-cm-long commercially available PBGF (NKT photonics, AIR 6-800) is used for this experiment. The inset of Fig. 1 shows an SEM image of the fiber cross section. The Rb vapor is confined to the core of the PBGF, which is 6-${\mu}$m in diameter as indicated in the figure. An external cavity diode laser (Toptica, TA 100) tuned to the $D_{1}$ line of $^{85}$Rb and set to a mode-hope-free scan serves as the weak probe. The power of the probe beam is maintained at $\sim$ 1 nW, which is below the saturation power of Rb vapor in the core of the PBGF \cite{Steck}. The CW output from a home-built Titanium sapphire (Ti:S) laser, tunable from 772 nm to 810 nm is used for Rb vapor generation. For pulsed measurements, the laser is mode-locked to generate 100-fs pulses at a repetition  rate of 80 MHz. Both the probe and vapor generation beams are free-space coupled into the fiber using 10X objectives. Including coupling losses, the transmission through the fiber exposed to Rb vapor gradually drops over a month to 23\% during continuous operation as compared to 70\%  for a pristine fiber. Formation of metallic clusters at the core walls make the transmission loss high over a broad frequency range. The transmission of a fully loaded fiber prior to vapor generation can be as low as 10\% and the transmission rises gradually to 23\% for off-resonant vapor generation as the Rb atoms are released from the core walls. The Rb vapor in the chamber is an isotopic mixture of $^{85}$Rb and $^{87}$Rb in a ratio of 7:3. As seen in Fig. 1, both  probe and vapor generation beams are sent counter propagating into the fiber, so that the weak probe can be easily separated. The probe beam after propagation is separated using a band-pass filter at 795 nm and is measured using a photodiode. 

\begin{figure}[hb]
\centerline{\includegraphics[width=0.5\columnwidth]{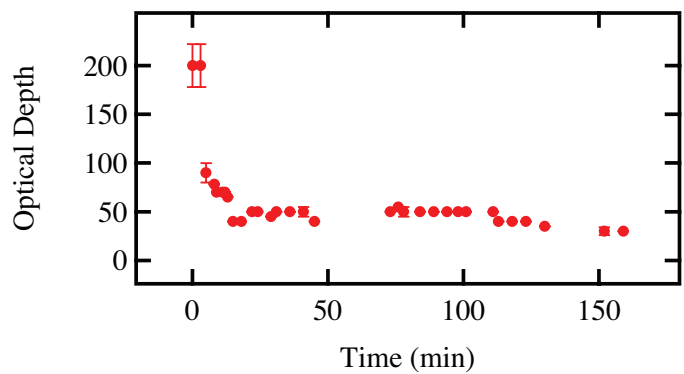}}
\caption{ Optical depth (OD) generated due to CW vapor generation is plotted as a function of elapsed time. It is observed that high OD's $\sim$ 200 are generated for a few minutes and thereafter the OD stabilizes to 50${\pm}$5 for over 100 minutes of  operation. The measurement is performed for 35 mW of vapor generation beam power transmitted through the PBGF.}
\end{figure}

We estimate the generated OD by fitting the absorption experienced by the probe to a Voigt function. Figure 2, shows the probe absorption (blue points) during a scan across the $D_{1}$ line of $^{85}$Rb and $^{87}$Rb at different  powers of the vapor generation beam. All detunings are measured with respect to the F = 3 $\rightarrow$ F' = 3 transition on the $^{85}$Rb $D_{1}$ line. We fit the probe absorption data to a Voigt profile (red curve) which takes into account both the homogenous (Lorentzian) and inhomogenously Doppler broadened (Gaussian) line shapes. Since the Rb atoms are confined to a 6-${\mu}$m core, the interaction time with the optical mode is limited to the time it takes for the atoms to move across the beam and leads to transit-time broadening of the absorption lines, which we take into account while simulating the Voigt function \cite{slepkov_optexp,slepkov_pra}.  The temperature of the Rb atoms inside the core is inferred to be $T$ $\sim$ 500 K from the Voigt function fit in Fig 2(a)-(b). The higher temperature of the atoms (500 K) as compared to the chamber temperature of 373 K appears to be a result of the high kinetic energy of the released atoms \cite{slepkov_optexp} which also results in the increased Doppler broadening of the Rb transition lines. We also note that the large OD's are only generated when the vapor generation beam is coupled into the fiber core. Secondly, the OD inside a 2-cm-long section of the chamber is 0.5, which tells us that the 0.5-mm-long air gap region between the fiber tip and the chamber view-port on each side contribute negligibly to the total OD when the vapor generation beam is coupled into the fiber. 

\begin{figure}[ht]
\centerline{\includegraphics[width=0.5\columnwidth]{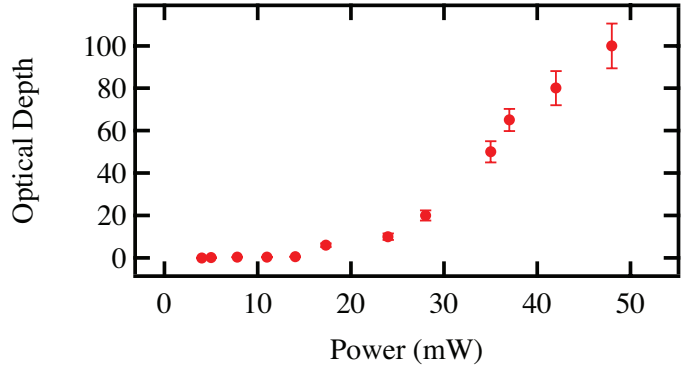}}
\caption{The generated OD is plotted as a function of power of the vapor generation beam transmitted through the PBGF. It is observed that at lower powers, the OD increases slowly. However, beyond a vapor generation beam power of 23 mW, the OD continues to increase.}
\end{figure}

We first demonstrate large OD's using a CW vapor generation beam. A strong CW beam at 805 nm is coupled into the fiber, and the generated OD is continuously monitored by measuring the transmission spectrum of the probe beam. Figure 3 shows the OD as a function of elapsed time. We observe that high OD's, as large as 200,  are generated for a few minutes and thereafter a stable OD = 50${\pm}$5  is observed for over 100 minutes,  which represents a 1000X improvement of operation time of the system over that previously reported \cite{Bhagwat, Saikat, Pablo, Viv1,Vivek_natphoton, Vivek_prl, Prathamesh, Walmsley}. For our Rb-PBGF system, an OD $\sim$ 50 corresponds to an atomic number density ${N{\approx}10^{19}}$ atoms/m${^{3}}$. The transmitted power of the vapor generation beam through the fiber is 35 mW for 150 mW of power at the input end. The low transmission through the PBGF is due to the presence of Rb at the core walls which increases scattering losses and modifies the guiding properties of the fiber over time after long exposure.
The generated OD is also measured as a function of vapor generation beam power as shown in Fig. 4. We see that beyond a  power of 23 mW, the generated OD continues to increase. This indicates that larger OD's can be generated using vapor generation beams with higher average powers. Currently we are limited by the CW power of the Ti:S oscillator, and we expect that higher OD's can be generated at larger average powers levels of the vapor generation beam.

\begin{figure}[ht]
\centerline{\includegraphics[width=0.5\columnwidth]{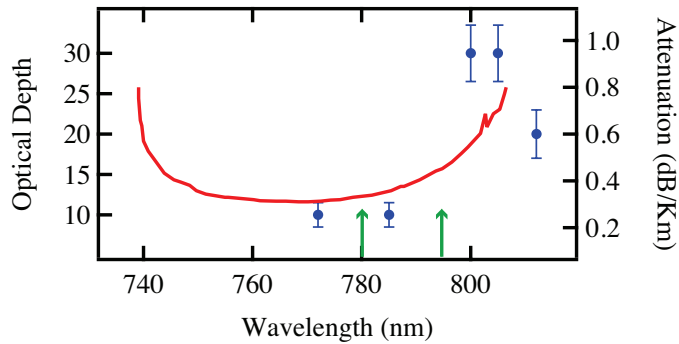}}
\caption{Optical depth (OD) generated due to the vapor generation beam is plotted as a function of its wavelength (blue points). The curve in red shows the transmission data of the fiber and defines its band-gap \cite{nkt}. The green arrows indicate the $D_{1}$ and $D_{2}$ transition of Rb. It is observed that the OD's generated at wavelengths within the band gap are lower than at wavelengths close to the edge and slightly beyond. As we move further away from the edge, the generated OD's decrease. }
\end{figure}

We also examine the wavelength dependence of the vapor generation beam within the band gap of the fiber. Figure 5 shows the generated OD as a function of wavelength of the vapor generation beam (blue points). The green arrows in the figure indicate the $D_{1}$ and $D_{2}$ transitions of Rb. The OD at each wavelength is measured for 30 mW of power transmitted through the fiber. We also plot the transmission data for the PBGF (red curve) \cite{nkt}. It is observed that the OD generated from 800 to 805 nm is 3X that generated at wavelengths within the band edge. We find that OD generation decreases as the wavelength of the vapor generation beam is tuned further away from the photonic band gap.  We believe this is due to increase in the guiding loss which reduces the power of the vapor generation beam as it propagates along the fiber. Thus, OD's are generated most efficiently when the vapor generation beam wavelength is chosen to be at the band edge of the PBGF. This is as expected since the optical mode would become more delocalized at the edge of the band-gap and more of the light from the vapor generation beam will be at the core walls. The Ti:S laser used for the experiment is tunable only to 772 nm, and hence we could not acquire data below this wavelength further into the bandgap.  

Typically for our experiments exploring nonlinear interactions with Rb vapor, we choose a vapor generation beam wavelength of 805 nm, since it is far away from both the Rb $D_{1}$ and $D_{2}$ lines and hence would not lead to any resonant interaction with Rb vapor or cause significant AC stark shifting of its hyperfine levels at high intensities \cite{slepkov_pra}. Additionally, it is  easy to separate a probe tuned to any Rb transition from the vapor generation beam using a spectral filter.

As a final test of the origin of the Rb vapor in the core, we performed the experiment using a mode-locked Ti:S laser. The pulsed source was used to study the impact of average versus peak power on the vapor generation process. A 100-fs pulse with a repetition rate of 80 MHz and average power of 100 mW (40 KW peak power) was used to generate the vapor.  Figure 6 plots the OD generated using a pulsed source, as a function of elapsed time. It is observed that a stable OD = 20${\pm}$3  is generated for over 100 mins at an average power of 22 mW  transmitted through the fiber. It can be inferred  that the OD's generated using a pulsed source are comparable to those using a CW beam at similar average powers as shown in Fig. 4.  This indicates that the density of Rb atoms inside the core of the PBGF depends on the average and not the peak power of the vapor generation beam.

\begin{figure}[ht]
\centerline{\includegraphics[width=0.5\columnwidth]{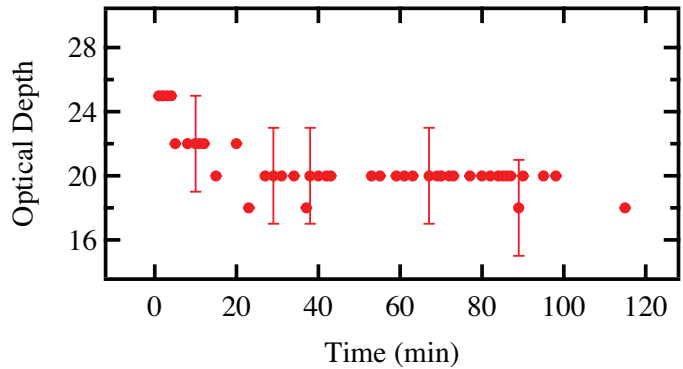}}
\caption{OD generated due to pulsed high peak power beam is plotted as a function of elapsed time. It is observed that a stable OD = 20${\pm}$3 is generated for over 90 minutes of operation. The average power through the core is 22 mW while the peak power is 40 KW.}
\end{figure}

Based on our observations, we propose a model that explains the dynamics of Rb atoms inside the fiber core. It has been suggested that the OD inside the core is generated by a fast mechanism and a slow mechanism \cite{Bhagwat}. The OD's generated by the fast process (i.e., laser desorption of Rb atoms chemically bonded to the core wall) are usually low (OD ${\sim}$ 0.5) but require short regeneration times ${\sim}$ 50 $\mu$s \cite{Bhagwat}, while those generated from the slow mechanism are extremely large (OD ${\sim}$ 1000), and last for only a few seconds \cite{slepkov_optexp}. The source for the generation of such high OD's by the slow mechanism are atoms that are contained in Rb nanoclusters which can be thermally evaporated at considerably low average power levels. As the average power of the vapor generation beam is low, the atoms cannot be prevented from sticking to the walls of the core and hence the vapor quickly disappears leading to operation time scales limited to a few seconds. However, atoms that are chemically bound to the surface of the core walls require larger powers to be released. As the power of the vapor generation beam is continuously increased, beyond those required for the slow mechanism, more and more atoms chemically held to the core walls are released  and a state of equilibrium is reached for each power level, leading to a stable and sustained OD for extended time scales ranging up to a few hours. Our results shown in Fig. 5, also indicate that the vapor generation efficiency is greater at wavelengths at the photonic band edge of the fiber. This is a result of more of the field being pushed into the cladding as the guiding efficiency lowers. Thus, as we tune the vapor generation beam away from the band gap, the intensity of the beam close to the walls of the fiber-core starts increasing which prevents the atoms from sticking to the walls, thereby assisting in the efficient generation of large OD's.

In summary we demonstrate continuous ODs for times that are ${\textgreater}$ 1000X longer than those  previously demonstrated. Our observations indicate that vapor generation is more efficient at the photonic band edge as the field mode inside the core is more concentrated near the surface which prevents the atoms from sticking to the core wall. We also establish that the vapor generation process depends on the average and not the peak power. We believe our system will allow us to attain continuous, strong light-matter interactions at the single-photon level.

We acknowledge financial support from the National Science Foundation (Grant no. PHY - 0969996) and DARPA via the QUINESS program.


\begin{thebibliography}{1}


\bibitem{Payeronel} T. Peyronel,	O. Firstenberg,	Q. Liang,	S. Hofferberth,	A. V. Gorshkov,	T. Pohl, M. D. Lukin, and V. Vuleti\'{c}, Nature \textbf{488}, 57 (2012).
\bibitem{Chen} W. Chen, K. M. Beck, R. B\"{u}cker, M. Gullans, M. D. Lukin, H. Tanji-Suzuki,	and V. Vuleti\'{c}, Science \textbf{341}, 768 (2013).
\bibitem{Tiecke} T. G. Tiecke, J. D. Thompson, N. P. de Leon, L. R. Liu, V. Vuleti\'{c}, and  M. D. Lukin, Nature \textbf{508}, 241 (2014).
\bibitem{Ritter} A. Reiserer, N. Kalb, G. Rempe, and S. Ritter, Nature \textbf{508}, 237 (2014).
\bibitem{Rauschenbeutel} J. Volz, M. Scheucher, C. Junge, and  A. Rauschenbeutel, Nat. Photonics \textbf{8}, 965 (2014).
\bibitem{Sven} N. K. Langford, S. Ramelow, R. Prevedel, W. J. Munro, G. J. Milburn, and  A. Zeilinger, Nature \textbf{478}, 360 (2011).
\bibitem{Imamoglu} H. Schmidt and A. Imamoglu, Opt. Lett \textbf{21}, 1936 (1996).

\bibitem{Kumar} P. Kumar, Opt. Lett \textbf{15}, 1476 (1990).
\bibitem{Bhagwat} A. R. Bhagwat, A. D. Slepkov, V. Venkataraman, P. Londero, and A. L. Gaeta, Phys. Rev. A \textbf{79}, 063809 (2009).
\bibitem{Saikat} S.Ghosh,  A. R. Bhagwat, C. K. Renshaw, S. Goh, and A. L. Gaeta, Phys. Rev. Lett \textbf{97}, 023603, (2006).
\bibitem{Pablo} P. Londero, V. Venkataraman, A. R. Bhagwat, A. D. Slepkov, and A. L. Gaeta, Phys. Rev. Lett. \textbf{103}, 043602 (2009).
\bibitem{Viv1} V. Venkataraman, P. Londero, A. R. Bhagwat, A. D. Slepkov, and A. L. Gaeta, Opt. Lett \textbf{35}, 2287 (2010).
\bibitem{Sprague} M. R. Sprague, D. G. England, A. Abdolvand, J. Nunn, X. Jin, W. S. Kolthammer, M. Barbieri, B. Rigal, P. S. Michelberger, T. F. M. Champion, P. St. J. Russell, and I. A. Walmsley, New J Phys \textbf{15}, 055013 (2013).
\bibitem{Epple} G. Epple,	K. S. Kleinbach,	T. G. Euser,	N. Y. Joly,	T. Pfau,	P. St. J. Russell, and R. L$\ddot{o}$w,  Nat. Commun \textbf{5}, 1 (2014).
\bibitem{Vivek_natphoton} V. Venkataraman, K. Saha and A. L. Gaeta, Nat. Photonics \textbf{7}, 138 (2013).
\bibitem{Luiten} C. Perrella, P. S. Light, J. D. Anstie, F. Benabid, T. M. Stace, A. G. White, and A. N. Luiten, Phys. Rev. A \textbf{88}, 013819 (2013).
\bibitem{Vivek_prl} V. Venkataraman, K. Saha, P. Londero, and A. L. Gaeta, Phys. Rev. Lett. \textbf{107}, 193902 (2011).
\bibitem{Prathamesh} P. S. Donvalkar, V. Venkataraman, S. Clemmen, K. Saha and A. L. Gaeta, Opt. Lett \textbf{39}, 1557 (2014).
\bibitem{Walmsley} M. R. Sprague, P. S. Michelberger, T. F. M. Champion, D. G. England,	 J. Nunn,	X.-M. Jin,	 W. S. Kolthammer,	 A. Abdolvand,	 P. St. J. Russell, and I. A. Walmsley, Nat. Photonics \textbf{8}, 287 (2014).
\bibitem{Blatt} F. Blatt, T. Halfmann, and T. Peters, Opt. Lett \textbf{39}, 446 (2014).
\bibitem{russel_hg} U. Vogl, C. Peuntinger, N. Y. Joly, P. St.J. Russell, C. Marquardt, and G. Leuchs, Opt. Exp \textbf{22}, 29375 (2014).
\bibitem{slepkov_pra} A. D. Slepkov, A. R. Bhagwat, V. Venkataraman, P. Londero, and A. L. Gaeta, Phys. Rev. A \textbf{81}, 053825 (2010).
\bibitem{Kasturi} K. Saha, V. Venkataraman, P. Londero, and A. L. Gaeta, Phys. Rev. A \textbf{83}, 033833 (2011).
\bibitem{Munro} W. J. Munro, K. Nemoto, and T. P. Spiller, New J Phys \textbf{7}, 137 (2005).
\bibitem{Steck} D. A. Steck, ``$^{87}Rb$ and $^{85}Rb$ $D_{1}$ line data," http://steck.us/alkalidata/ (2013).
\bibitem{slepkov_optexp} A. D. Slepkov, A. R. Bhagwat, V. Venkataraman, P. Londero, and A. L. Gaeta, Opt. Exp \textbf{16}, 18976 (2008).
\bibitem{nkt} "NKT Photonics PCF," http://assets.newport.com/webDocuments-EN/images/15128.PDF
\end{thebibliography}
\end{document}